\documentclass{article}{\twocolumn}
\usepackage{graphicx} 
\usepackage{hyperref}
\usepackage{authblk}
\usepackage[a4paper, total={6in, 8in}]{geometry}
\title{Development of a 3D-printed canine head phantom for veterinary radiotherapy}
\author[1,2]{Sandhya Rottoo}
\author[3]{Luke Frangella}
\author[2]{Magdalena Bazalova-Carter}
\author[2]{Olivia Masella}
\affil[1]{McGill University, Department of Physics}
\affil[2]{University of Victoria, Department of Physics and Astronomy}
\affil[3]{Proto3000}
\date{\today}
\usepackage{subcaption}
\usepackage{float}
\usepackage{gensymb}

\begin{document}
\sloppy

\maketitle

\begin{abstract}

\textbf{Purpose:} To develop the Ultimate Phantom Dog for Orthovoltage Glioma Treatment (UPDOG), an anatomically-correct phantom which mimics a dog's head, for quality assurance (QA) of kilovoltage (kV) radiotherapy treatments.

\textbf{Methods:} A computed tomography (CT) scan of a canine glioma patient was segmented into bone and soft tissue using 3DSlicer. The segments were converted to stereolithographic (STL) files and smoothed in Fusion360. A slit to accommodate a radiochromic film (RCF) was added at the location of the glioma tumor. UPDOG was 3D printed on a polyjet printer using VeroUltraWhite ($\rho$ = 1.19-1.20 g/cm\textsuperscript{3}) for the bone and Agilus30 ($\rho$ = 1.14-1.15 g/cm\textsuperscript{3}) for the soft tissue. CT scans of UPDOG were acquired on a clinical CT scanner. An LD-V1 RCF was inserted into UPDOG and irradiated with a kV x-ray source from two angles. The delivered dose to the RCF was compared to Monte Carlo (MC) simulations performed in TOPAS.

\textbf{Results:} The bone and soft tissue segments in UPDOG were mimicked the patient anatomy well with tube voltage-dependent CT numbers.  The contrast in HU was of 49, 47 and 50 HU for the 80, 100, and 120 kVp scans, respectively, sufficient for anatomy visualization. The irradiations delivered a maximum dose to RCF of 284 mGy which was compared to the results of MC simulations using a depth dose curve and central-axis (CAX) beam profiles. The mean difference in CAX profiles and PDD between RCF and MC results was 15.9\% and 2.3\%, respectively.  

\textbf{Conclusions: } We have demonstrated that UPDOG is a useful QA tool for kV canine radiotherapy. UPDOG successfully anatomically mimicked the dog anatomy, with a reduced but sufficient bone contrast. We showed that dose delivered to a canine glioma with kV x-rays can be successfully measured with an RCF positioned at the tumor location.

\end{abstract}

\section{Introduction}


Anatomically correct anthropomorphic phantoms are a useful tool for quality assurance in radiotherapy and imaging. They allow for repeatable, accurate dosimetry measurements that mimic real treatments, and are forgiving of repeated exposures to radiation, making them ideal for validating new treatment systems and methods. 

3D printing has emerged as a cost-effective and efficient solution to easily manufacture anatomically-realistic phantoms \cite{3dreview} \cite{3dreview2}. This technique is especially popular in the context of small-animal dosimetry \cite{mouse},\cite{Bentz2016} where costs are scaled down by the size of the animal, but it has also been used to manufacture a water-equivalent dog skull \cite{tanabe2023}, various parts of humans \cite{Ehler_2014} \cite{gear2016} and even moving phantoms mimicking anatomical motion \cite{Laidlaw2023} \cite{Mayer2015}. The versatility of 3D printing techniques allows for specific, tailored solutions to a study's needs.

The three main types of 3D printing are fused deposition modeling (FDM), where a thermoplastic is melted and cooled layer by layer onto a build plate; selective laser sintering (SLS), where a laser sinters a polymer powder into a 3D shape; and stereolithography (SLA), where an ultraviolet laser selectively cures a resin layer by layer onto an inverted build plate. Quality and cost of a print can vary dramatically depending on which technique and materials are used. 

PolyJet printers use a technique which combines FDM and SLA to build parts from multiple materials simultaneously. Multiple nozzles each extrude a different type of resin in patterns, which are then selectively cured by an ultraviolet laser. This process is repeated layer-by-layer. Supports are printed with a water-soluble material to hold overhangs and can be easily removed after the printing process. More sophisticated printers can even create heterogeneities that mimic different anatomical structures. This makes it the technology of choice for anatomical models since multiple materials can be used to create contrast between different organs and tissues. For use in x-ray imaging, the materials ideally also have the same radiographic properties as the tissue they represent. This allows for the dose distribution within the phantom to mimic what it would be on the living specimen \cite{esplen2019preclinical}. In reality, the material properties will not match the real tissues exactly since they need to be manufactured in such a way that they can be printed with a PolyJet printer. In order to use a phantom for radiotherapy end-to-end (E2E) quality assurance (QA) testing, it is common to place radiochromic films (RCFs) in suitable areas within a phantom to measure radiotherapy dose distributions \cite{esplen2019preclinical} \cite{steinmann2020mrigrt} \cite{CUNHA2016223}. 

This work focuses on the development of the Ultimate Phantom Dog for Orthovoltage Glioma treatment (UPDOG), a 3D printed realistic E2E dog head phantom. Computed Tomography (CT) scans of a canine brain cancer patient were segmented into bone and soft tissue segments and 3D printed with two materials to form UPDOG. A slit in the brain of UPDOG to hold an LD-V1 RCF  included. UPDOG was irradiated with an x-ray source and film dose was compared to dose calculated by Monte Carlo methods.

\section{Method}

\subsection{Development of UPDOG}

\begin{figure}[hbt!]
    \centering
    \includegraphics[width=\linewidth]{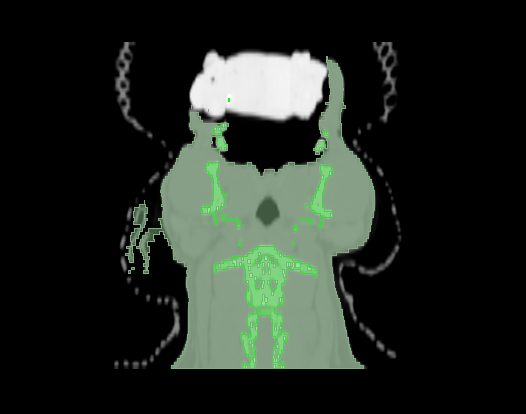}
    \caption{A slice of the CT scan of the canine patient, with the bone (green) and soft tissue (gray) segments, contoured in 3DSlicer. The bite block, which has a similar density to bone, is shown in white. The immobilization thermoplastic mask seen around the dog patient was not included in phantom design.}
    \label{fig:segmentation}
\end{figure}

This study used the Kilovoltage Optimized AcceLerator Adaptive therapy (KOALA) x-ray system at the University of Victoria \cite{oconnell2024}, which is being developed with the aim to treat canine cancer patients with kilovoltage (kV) x-ray beams. The phantom needed to mimic a dog structurally and in radiographic properties, as well as incorporate some method of measuring dose delivery. It is necessary to validate KOALA's ability to effectively treat deep-seated tumours, so a glioma patient was used as a model for UPDOG. Computed tomography (CT) scans of the patient were obtained from a veterinary clinic and segmented using Python and 3DSlicer (version 5.6.2, Brigham and Women’s Hospital, Boston, MA) \cite{3DSlicer}. The bite block, thermoplastic mask, and breathing tube used to acquire the CT scans were manually removed, and the image was separated into two segments of bone and soft tissue  (Figure  \ref{fig:segmentation}). The segments were converted into stereolithographic (STL) files in 3DSlicer and smoothed using Fusion360 (version 16.9.0.2204, Autodesk, Mill Valley, CA). A $3\times6\times0.07$ cm slit was added into the middle of UPDOG's head at the location of the glioma tumor to accommodate an RCF.  

\begin{figure}[hbt!]
    \centering
    \includegraphics[width=\linewidth]{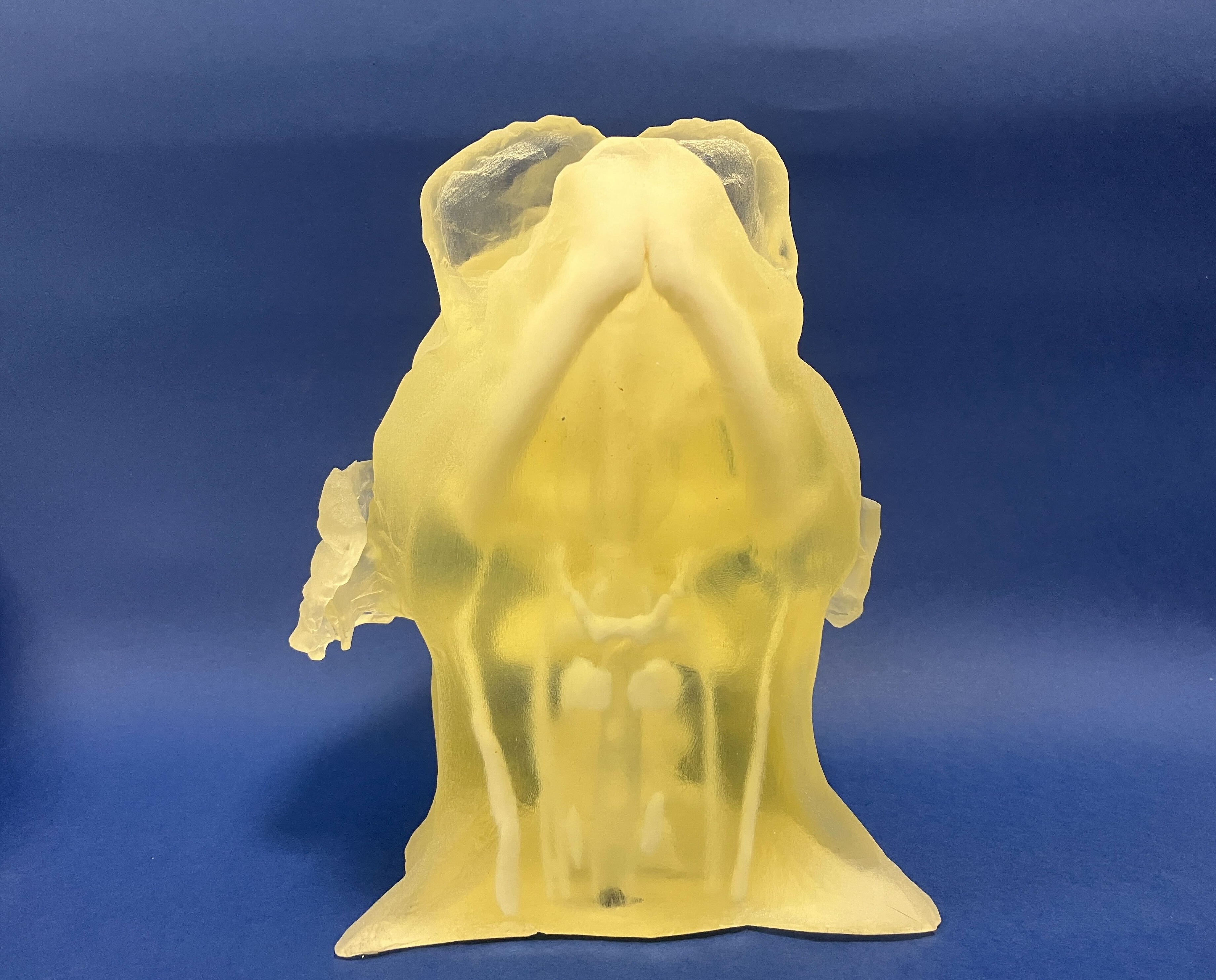}
    \caption{UPDOG, 3D printed with PolyJet technology. The bone (white) and soft tissue (clear) are seen as distinct segments, as contoured in Figure \ref{fig:segmentation}.}
    \label{fig:updog}
\end{figure}

The phantom was printed on a Stratasys Polyjet J850 \cite{printer} by Proto3000, Inc. (Vaughn, ON) and it is shown in Figure \ref{fig:updog}. The bone was made of VeroUltraWhite \cite{veroUltraWhite} ($\rho$ = 1.19-1.20 g/cm\textsuperscript{3}) and the soft tissue was made of Agilus30 \cite{Agilus30} ($\rho$ = 1.14-1.15 g/cm\textsuperscript{3}). The materials were selected based on density and reported Hounsfield Unit (HU) values \cite{materialsCT}. Though the HU value of VeroUltraWhite does not have a literature value, it is part of the Vero\texttrademark family and has the same properties in the polymerized state as VeroWhitePlus with a higher density. Therefore HU values slightly higher than what was measured for VeroWhitePlus were expected.  The total cost for commercial printing of UPDOG was \$2,875 CAD, most of which was attributed to the selected materials. 

\subsection{Phantom CT imaging}

A CT scan of UPDOG was taken with an Optima CT580 scanner (GE Healthcare, Chicago, Il) with tube voltages 120, 80 and 100 kVp, a tube current-time product of 250 mAs and the clinical brain scan protocol. CT images of the phantom and dog patient were compared qualitatively and quantitatively. For each CT scan, the mean HU values of the bone and soft tissue segments were measured using 3DSlicer.

\subsection{Phantom irradiations}

\begin{figure*}[htbp]
    \centering
    \begin{subfigure}[b]{0.45\textwidth}
        \centering
        \includegraphics[width=\textwidth]{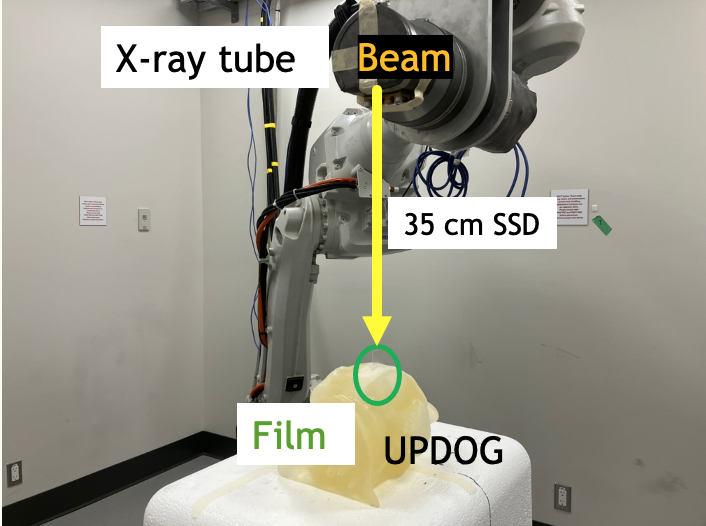}
        \caption{0\degree irradiation}
        \label{fig:figure1}
    \end{subfigure}
    \hfill
    \begin{subfigure}[b]{0.45\textwidth}
        \centering
        \includegraphics[width=\textwidth]{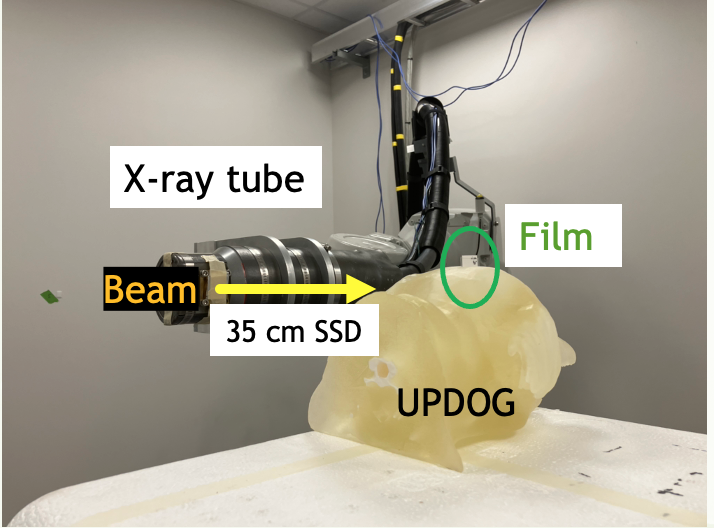}
        \caption{90\degree irradiation}
        \label{fig:figure2}
    \end{subfigure}
    \caption{Pictures of KOALA treating UPDOG with a 225 kVp beam, collimated to 1 mm, filtered with 2 mm of aluminum from an SSD of 35 cm. (a) shows the beam at an angle of 0\degree and (b) shows the beam at an angle of 90\degree.}
    \label{fig:treatment-setup}
\end{figure*}

A low-energy low-dose LD-V1 RCF (Ashland, Bridgewater, NJ) was used for UPDOG dose measurements. A  $3\times6$ cm piece of RCF was placed into a plastic pocket to prevent scratching or coating in residue from the soft Agilus30, and inserted into the UPDOG brain slit. Irradiations were delivered with a 225 kVp x-ray source (MXR 225/26 by Comet, Flamatt, Switzerland) affixed to one of two IRB 4600 Robotic Arms (ABB Ltd., Zurich, Switzerland). The beam was collimated  to 1 mm with a 2-mm thick tungsten collimator at 3.5 cm from the x-ray tube focal spot and filtered with 2 mm of aluminum, with a source-to-surface distance (SSD) of 35cm.
Two beam angles were delivered to two films: from a 0\degree angle (x-ray tube vertical) for an exposure time of 40s and from a 90\degree angle (x-ray tube horizontal) for an exposure time of 120 s, both with a tube current of 2.5 mA (Figure \ref{fig:treatment-setup}).

The dose-to-film was measured by means of a calibration curve. The data for this curve \cite{masella2024} was obtained by delivering known doses between 5-500 mGy to pieces of LD-V1 film from the same batch and beam quality (225 kVp, 2 mm Al filtration), calculating the net optical density (netOD) for each dose level: 

\begin{equation}
    netOD = log_{10}\left(\frac{I_{flood}}{I}\right) - log_{10}\left(\frac{I_{flood}}{I_0}\right),
\end{equation}

where $I$ and $I_0$ are pixel values of irradiated and unirradiated film, and $I_{flood}$ is the pixel value of the flood field of the corresponding region-of-interest. Films were scanned at 150 dpi resolution and the  red channel was used to create the calibration curve that was fitted with a rational function.

\subsection{Monte Carlo simulations}

TOPAS (TOol for PArticle Simulation), released in 2014, \cite{TOPAS} \cite{TOPAS2} is a medical physics-oriented wrapper around Geant4 \cite{ALLISON2016186} \cite{1610988} \cite{AGOSTINELLI2003250}, a popular Monte Carlo (MC) based particle transport simulation framework which has been extensively validated against experimental data \cite{AMAKO200644}. TOPAS can build simulation geometry based on STL files with user-defined materials, as well as score dose in any geometry components. TOPAS can also implement low-energy transport physics with modules like Livermore \cite{CHAMPION2009745}, which was used for this study. 

For the simulated irradiations, the bone and soft tissue STL files were simplified to have less than 10,000 triangles and imported into TOPAS. Since the materials used for printing are patented, both were given the material properties of PMMA resin with correct material densities. The LD-V1 RCF in the phantom slit was simulated as a 239-um thick water slab in the middle of UPDOG's head, with a 25 $\mu$m central active layer used for scoring dose-to-medium. The film dose was scored in $22\times 41$ mm voxels, and standard deviation was scored for uncertainty.  The 225 kVp beam filtered with 2-mm aluminum was simulated in EGSnrc \cite{egsnrc} and collimated to 1 mm using a 2-mm thick tungsten collimator at 35 mm from the source, mimicking the true irradiation geometry. A phase-space file was then generated and scored to use in TOPAS. The beam was simulated with 5.5$\times10^6$ original histories that were recycled 60 times. The simulations were performed using the Digital Research Alliance of Canada's cluster with 50 CPUs requested and took around 3h to complete.

\section{Results}

\subsection{ UPDOG CT images}

\begin{figure}[hbt!]
    \centering
    \begin{subfigure}[t]{0.49\columnwidth}
        \centering
        \includegraphics[width=\textwidth]{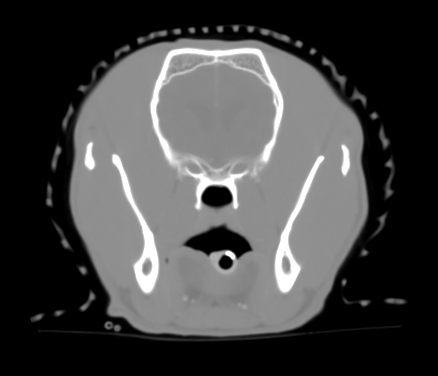}
        \caption{Dog patient (W = 1300,L = 140)}
         \label{fig:dog scan}
    \end{subfigure}
    \hfill
    \begin{subfigure}[t]{0.49\columnwidth}
        \centering
        \includegraphics[width=\textwidth]{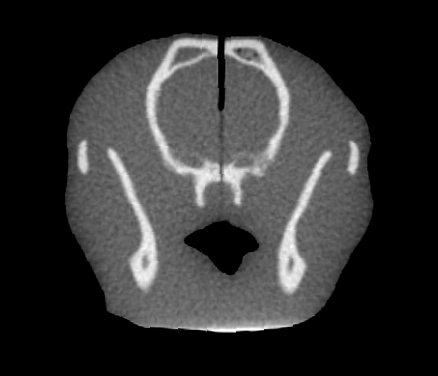}
        
        \caption{UPDOG (W = 200, L = 200)}
        \label{fig:updog scan}
    \end{subfigure}

\caption{A  CT image of (a) the original canine patient, imaged at 120 kVp and 210 mAs and (b) of UPDOG, imaged at 120 kVp and 250 mAs. The window and level of each image are indicated in the subcaptions}
\label{fig:scans}
\end{figure}

Figure \ref{fig:scans} compares  CT images of the original dog patient with the CT scan of UPDOG.  Deliberate changes are visible, including the removal of the breathing tube, inclusion of the RCF slit, and the nasal cavity being filled with soft tissue instead of being hollowed since supports in this cavity could not be removed.

For each CT scan, the mean HU value of the bone and soft tissue segments were measured using 3DSlicer. (Table \ref{table:HU}). The HU values for the soft tissue  (Agilus30) ranged from 72 to 84, following a similar trend to the literature range \cite{materialsCT}. The bone (VeroUltraWhite) had higher HU values than what was measured by Neumann et al. \cite{materialsCT} for VeroWhitePlus. Though the HU values of both materials expectedly changed with tube voltage, the difference between them stays relatively constant, allowing for a good contrast at all three tube voltages. For the real dog patient which was scanned at 120 kVp, the mean HU values were of 785 and 96 for the bone and soft tissue respectively, clearly a much larger contrast than in UPDOG.

\begin{table}[hbt!]
\begin{footnotesize}
\centering
\begin{tabular}{|l|ll|l|}
\hline
  Tube voltage     & \multicolumn{2}{l|}{Mean CT number (HU)} & Difference   \\ \cline{2-3}
 (kVp) & \multicolumn{1}{l|}{Bone}  & Soft tissue  & in HU \\ \hline
80  & \multicolumn{1}{l|}{121}      &     72   &   49  \\ \hline
100 & \multicolumn{1}{l|}{127} & 80    &  47   \\ \hline
120 & \multicolumn{1}{l|}{134}      &     84    & 50   \\ \hline
\end{tabular}
\caption{The mean CT numbers of the bone (VeroUltraWhite) and soft tissue (Agilus30) in UPDOG, measured at a tube current of 250 mAs and varying tube voltages. The difference in HU between the materials is also shown.}
\label{table:HU}
\end{footnotesize}
\end{table}

\subsection{UPDOG irradiation}

Two irradiations were performed with the setups shown in Figure \ref{fig:treatment-setup}, and the dose to film was calculated and compared to TOPAS MC simulations. 

\begin{figure}[hbt!]
    \centering
    \includegraphics[width=0.99\linewidth]{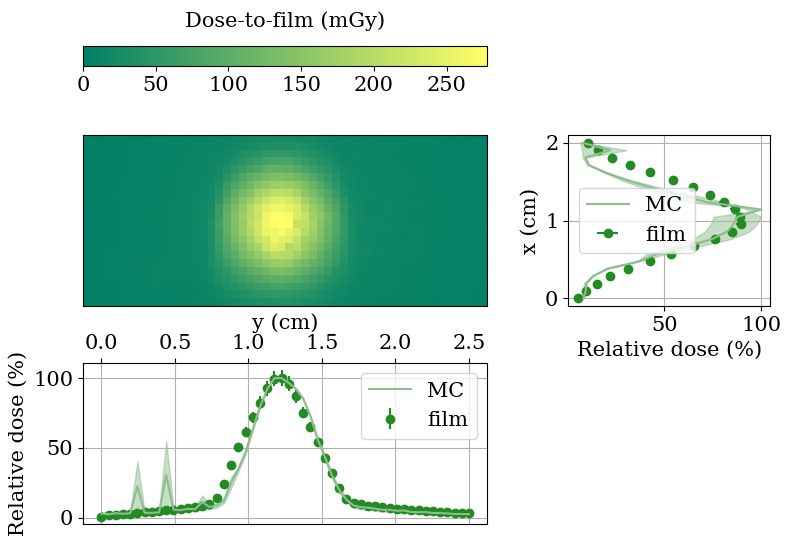}
    \caption{LD-V1 film dose distribution for the 90\degree irradiation, and the associated 1-D dose profiles compared to a normalized TOPAS Monte Carlo simulation dose, with associated uncertainties. The dose profiles were calculating by averaging all pixel values along each row or column. }
    \label{fig:dose-perp}
\end{figure}

For the 90\degree irradiation, the dose was delivered to 1 cm below the top of UPDOG's head. Figure \ref{fig:dose-perp} shows that the maximum dose delivered to the LD-V1 RCF at the centre of the beam was 266 mGy. The simulated dose was normalized to the film at the centre of the beam. 

\begin{figure}[hbt!]
    \centering
    \includegraphics[width=0.99\linewidth]{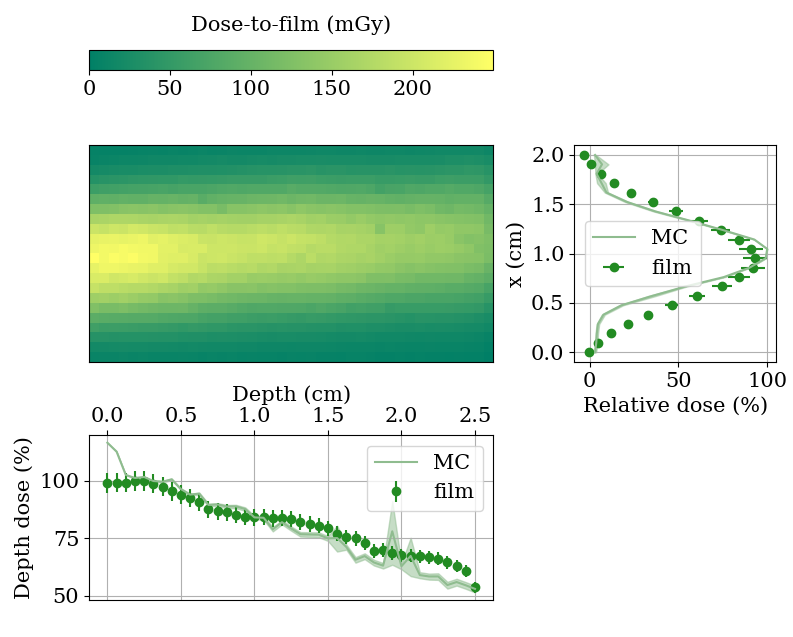}
    \caption{The film dose distribution for the 0\degree treatment, and the associated 1-D dose profiles compared to a normalized TOPAS Monte Carlo simulation dose, with associated uncertainties. The dose profiles were calculating by averaging all pixel values along each row or column. }
    \label{fig:dose-parallel}
\end{figure}

For the 0\degree irradiation, Figure \ref{fig:dose-parallel} shows that the maximum dose delivered occurred at the surface of UPDOG and was of 249 mGy. This is lower than for the 90\degree irradiation due to the lower exposure time and shallower beam penetration. The simulated depth-dose curve was normalized to the film at a depth of 1 cm.

\section{Discussion}

It is evident that UPDOG replicates the original dog anatomy with excellent accuracy and resolution, but the CT contrast between bone and soft tissue is not as high. This is due to the small material difference limited by the requirements of PolyJet printers. 

The experimental dose distributions match the simulated ones well once normalized, but some discrepancies are visible in the 0\degree irradiation. This could be due to several factors. First, the materials used to fabricate UPDOG are patented, so their simulated atomic composition is unknown and likely not accurate. This could affect the amount of x-ray and electron interactions that take place in the simulated UPDOG. Additionally, the resolution of PolyJet prints is of 0.1-0.3mm, which could lead to discrepancies between the STLs of UPDOG used in simulation and the actual UPDOG.

Typical canine brain radiation therapy involves a dose of 2-5 Gy/fraction, for multiple fractions spread out over many days \cite{dogsurvival}. The dose delivered to an LD-V1 by the KOALA x-ray tube scales linearly with time, so a 266 s ($\sim$ 5 min) irradiation would be required to achieve a 2 Gy dose-to-tumor for the 90\degree beam for a more realistic E2E QA test.

In the future, KOALA will deliver modulated collimated coplanar and non-coplanar arc treatments that create complex dose distribution maximizing the dose-to-tumor while minimizing dose-to-tissue, which would increase exposure time. These irradiations are beyond the scope of this work and will be explored in future studies.

\section{Conclusions}

UPDOG, a realistic phantom based on computed tomgraphy images of a canine glioma patient which incorporates a slit for a radiochromic film, was developed. UPDOG was 3D printed with PolyJet technology using Agilus30 for the soft tissue and VeroUltraWhite for the bone. UPDOG CT images revealed anatomical resemblance to the dog patient, albeit the bone contrast was decreased. Two proof-of-concept x-ray tube irradiations were delivered using the dual-robot KOALA system that demonstrated the utility of UPDOG for dose verification purposes. UPDOG will be used in the future for KOALA end-to-end quality assurance testing.

\section{Acknowledgements} 

We would like to thank the Natural Sciences and Engineering Research Council of Canada (NSERC) for providing us with USRA and Discovery Grant funding and the Canada Research Chairs program. The construction of the phantom was partially funded by Linden Technologies, Inc. This research was enabled in part by support provided by Calcul Québec (calculquebec.ca) and the Digital Research Alliance of Canada (alliance.can.ca). We would like to thank Nathan Clements for help with the naming of UPDOG, Michael Weil for his expertise on canine anatomy and Eric Lindsay from VCA Canada Central Victoria Veterinary Hospital for providing the patient CT data. 

\section{Data availability statement}

The CT scans and STL files of UPDOG, and the Monte Carlo parameter and output files are made publicly available on the XCITE lab \href{https://web.uvic.ca/~bazalova/UPDOG/}{webpage}. 

\section{Conflicts of interest}
Luke Frangella is an employee of Proto3000, who printed UPDOG. The other authors have nothing to disclose.
\bibliography{bib}
\bibliographystyle{unsrt}

\end{document}